\documentclass{article}

\usepackage{arxiv}

\usepackage[utf8]{inputenc} % allow utf-8 input
\usepackage[T1]{fontenc}    % use 8-bit T1 fonts
\usepackage{hyperref}       % hyperlinks
\usepackage{url}            % simple URL typesetting
\usepackage{booktabs}       % professional-quality tables
\usepackage{amsfonts}       % blackboard math symbols
\usepackage{nicefrac}       % compact symbols for 1/2, etc.
\usepackage{microtype}      % microtypography
\usepackage{lipsum}
\usepackage{graphicx}
\usepackage{wrapfig}
\usepackage{amsmath}

\DeclareMathOperator*{\argmin}{arg\,min}

\title{Fairness for Whom?
Critically reframing fairness with Nash Welfare Product
}

\author{
  Ansh Patel\thanks{author works at Electrophysiology, Memory and Navigation Lab at Columbia University and is a prospective graduate student in CS conducted this research in independent capacity.} \\
  \\
  Columbia University\\
  New York, NY 10027 \\
  \texttt{ansh.patel@columbia.edu} \\
}

\begin{document}
\maketitle

\begin{abstract}
Recent studies on disparate impact in machine learning applications have sparked a debate around the concept of fairness along with attempts to formalize its different criteria. Many of these approaches focus on reducing prediction errors while maximizing sole utility of the institution.
This work seeks to reconceptualize and critically frame the existing discourse on fairness by underlining the implicit biases embedded in common understandings of fairness in the literature and how they contrast with its corresponding economic and legal definitions. This paper expands the concept of utility and fairness by bringing in concepts from established literature in welfare economics and game theory. We then translate these concepts for the algorithmic prediction domain by defining a formalization of Nash Welfare Product that seeks to expand utility by collapsing that of the institution using the prediction tool and the individual subject to the prediction into one function. 
We then apply a modulating function that makes the fairness and welfare trade-offs explicit based on designated policy goals and then apply it to a temporal model to take into account the effects of decisions beyond the scope of one-shot predictions. We apply this on a binary classification problem and present results of a multi-epoch simulation based on the UCI Adult Income dataset and a test case analysis of the ProPublica recidivism dataset that show that expanding the concept of utility results in a fairer distribution correcting for the embedded biases in the dataset without sacrificing the classifier accuracy.

\end{abstract}

% keywords can be removed

\section{Introduction}
The concept of fairness has gained a lot of interest in the context of machine learning research in light of several high-profile studies (\cite{angwin2016machine}; \cite{buolamwini2018gender}) which demonstrated that real-world applications exhibited discriminatory behavior in their decisions. The subsequent criticism of opaque algorithmic decision-making models used to train such applications have resulted in a lot of work and discussion around achieving fairer outcomes.

A significant body of work by computer researchers have focused on introducing mathematical formulations for a variety of fairness criteria followed by different approaches that seek to satisfy them. Two of the commonly discussed fairness criteria, namely classification parity and calibration seek to provide concrete constraints that can be applied on predictions to minimize disparate impact effects on protected groups like race or gender The oft-discussed applications of the such fairness criteria either seek to ensure some sort of parity among the true and false positive rates -- namely demographic parity (\cite{feldman2015certifying}) and equalized odds (\cite{hardt2016equality}) or seek to ensure that the outcomes are independent of the protected group attributes (\cite{pleiss2017fairness}). However, recent work (\cite{corbett2018measure}) have underlined several criticisms and statistical limitations as well as incompatibilities (\cite{berk2017fairness}) of such criteria and their applications. In particular, such criteria have been shown to be blind to the underlying distributions of risk and prone to the problem of infra-marginality. In addition, adjusting for those underlying differences have been demonstrated to only exacerbate biases and increase misclassification risks.

Fairness criteria that seek to address such limitations (\cite{corbett2017algorithmic}) often focus on utility of an algorithmic decision that’s ultimately one-shot in nature, which in turn exposes a larger issue that plagues most existing criteria and its associated discourse. Namely, they make assumptions about the fairness of decisions that are seemingly isolated in a vacuum, and whose utility is measured almost exclusively from the narrow perspective of the institution or the user of the prediction tool. For instance, in a simple loan-lending scenario, the scope of most fairness criteria is limited to the moment the prediction is made with little interest in modeling how that decision affects the criteria it seeks to uphold or the larger distribution. In the same example, fairness is often held up against accuracy with the cost of misclassification being seen purely as a cost of bank profit with little regard to effects on the prospective lender who was misclassified.

In this paper, we begin with a simple question of “fairness for whom?” by claiming that the mathematical formalizations of fairness criteria proposed by computer researchers are embedded with their own biases and are often at odds with the fairness they seek to achieve. We extend arguments made by recent work (\cite{corbett2018measure};\cite{kleinberg2018algorithmic}) that cite the limitations of achieving fairness when biases are “baked in” the dataset. We believe that computer researchers don’t have to reinvent the wheel when it comes to formalizing fairness. Instead, they can rely on foundations laid by economic and legal definitions of fairness as well as their formalizations which have decades of literature behind them. 

Our goal in this paper is to outline a brief critique explaining the limitations of the dominant interpretation of fairness in the field and describe ways to reconceptualize them by bringing in concepts from welfare economics and game theory. In particular, we do this by rethinking fairness as a joint function of societal utility, thus encompassing both the institution using the prediction tool and the individual(s) that are subject to it, while seeking to  minimize individual disparate impact. For this purpose, we adapt Nash Welfare Product, a popular mainstay for decades in welfare economics and agent-based game theory, which binds the utility of every member of the society including the institution into a combined product and seeks to push all towards an equilibria. We apply it to a binary classification problem and through results of our analysis discuss how achieving fairness doesn’t have to be seen as a “cost” that needs to be balanced. By bringing in a wealth of concepts from other fields, we wish to treat fairness from an interdisciplinary lens and propose a possible step forward.

\section{Background}
Significant body of work proposing applications of the aforementioned fairness criteria work have focused on minimizing predictive error rates like false positives and false negatives while maintaining the accuracy of the classifier by introducing their own interventions into the machine-learning pipeline. The approaches within this framework differ based on what point of the pipeline the intervention is introduced, namely – preprocessing, algorithmic and post-processing (\cite{friedler2018comparative}). There are embedded philosophical assumptions made with each of those approaches, even if they all seek to ultimately achieve fairer decisions. For instance, pre-processing approaches accept that biases are embedded in the training dataset and seek to provide corrective measures to account for that influence on the prediction tools. On the other hand, post-processing approaches believe that given the barriers of interpretability and transparency of the such prediction tools, making adjustments after the fact to minimize disparate impact on any particular group is a preferable approach. Our paper will be exclusively focusing on post-processing approaches in both our criticism and our proposed reconceptualization. As mentioned earlier, some of the existing criticism laid out by \cite{corbett2018measure} exposes significant limitations of existing fairness approaches. In particular, they find differing risk distributions and infra-marginality as some of the key limitations that blindside the fairness criteria when they are applied in practice.

Fairness has been a concern not just for computer researchers but for economic and legal scholars for many decades before. There’s no dearth of literature that seeks to expand upon the many aspects of fairness as well as conceptual incompatibilities of it when seen in context of legal effects of a proposed policy (\cite{kaplow2009fairness}). In particular, welfare economics and game theory have amassed substantial literature that is interested in fairness. While in both fields, fairness is usually seen from the context of distribution of divisible resources and ensuring that the resulting condition is Pareto-optimality, or in other words nobody is worse off because of the decision; we can adapt some of the concepts for purposes of thinking about fairness of prediction tools in machine learning. In particular, recent work that thinks about the utility (\cite{corbett2018measure}) or temporally models the long-term effects of a prediction (\cite{liu2018delayed}) are a promising step in existing research towards this direction. But often, as explained in \cite{kleinberg2018algorithmic}, fairness is insufficient to overcome the “baked in” biases that are embedded in the dataset based on historical and existing practices. For instance, applying existing fairness constraints on whether someone will recidivate or not is insufficient to fully capture the differences in risk distribution let alone provide any mechanism to correct those biases beyond the scope of one-shot prediction. 

To expand further on work in this direction, we adapt Nash Welfare Product (NWP henceforth) which has been subject of much discussion in welfare economics and game theory (\cite{kaneko1979nash}). In particular, NWP’s performance in context of fairness has received particular attention (\cite{caragiannis2016unreasonable}; \cite{venkatasubramanian2018much}) in recent years. One recurring reason often cited is that its multiplicative nature allows welfare to be more distributive than a simple summation would allow (see the Social Planner Problem, for instance as described by \cite{hu2018welfare}). We believe that a summation as suggested by \cite{hu2018welfare} would conceal the underlying distribution differences. Our purpose of utilizing NWP is two-fold; to expand thinking about utility of an algorithmic prediction beyond that of its immediate user by modeling its effects on wider society; and to create a simple one-step feedback model that adjusts weights of individuals based on both welfare and fairness constraints. 

Traditionally, Nash Welfare has conceptually been seen as a halfway solution between utilitarian models and Rawlsian welfare models.. This is in line with our stated objectives which allows us to combine the utility of both the institution by minimizing loss in prediction accuracy as well as maximizing utility of the individuals and groups who are subject to the prediction. In the case of algorithmic decisions, it would mean a prediction that is seen as fair to both the institution and to the subject of the prediction. In the example of loan lending scenario, this would result in making explicit the potential trade-offs between the profits of the bank and the potential gains and losses of the individual seeking a loan. Thus by taking into account the effects of a prediction on both entities that have stake in the decision and then shifting weights for future distributions based on joint welfare and fairness constraints, NWP offers a good starting point for the objectives of this paper.

\section{Recent Work}

The criticisms and limitations of the dominant fairness criteria has resulted in recent work from scholars who have sought to expand the scope of thinking about fairness and applying it in context of algorithmic predictions. In particular, a formalization of the Social Welfare Planner problem (\cite{hu2018welfare}) and applying it on a binary classification problem is the nearest parallel to our paper’s goals. \cite{hu2018welfare} work proposes a weighted summation of each individual’s utility and seeks to maximize that while minimizing the loss on their linear SVM. This builds on the authors’ earlier work (\cite{hu2017short}) on interventions in labor market. On a similar front, other scholars (\cite{kleinberg2018algorithmic}) have also advocated for bringing concepts from welfare economics to expand inquiry into algorithmic fairness. \cite{kleinberg2018algorithmic} also utilize the Social Planner albeit in a different way by finding an equitable distribution of the privileged and discriminated group that would maximize the utility of the institution, in their case a college considering admissions. The work by \cite{hu2018welfare} and \cite{kleinberg2018algorithmic} is the clearest adjacent work to our efforts.

In addition to that, the aforementioned work to create a temporal model (\cite{liu2018delayed}) offers the first concrete example that seeks to model effects of one-shot prediction on simulations of FICO credit scores. \cite{liu2018delayed}’s paper still views its simulated utility from the narrow context of bank’s utility or profit, but their contributions in creating a feedback-response model that takes long-term effects of a one-shot prediction into consideration are an important step in moving away from the paradigm of looking at a decision in a vacuum.

\section{Problem Definition}

We will first formalize the Nash Welfare Product (NWP) as it is often used in agent-based simulations in game theory and then adapt it for our binary classification problem.
Consider a society with i individuals whose utility is a function of their initial income $x_{i}$ and their proposed outcome $y_{i}$ such that their individual utility can be calculated as $U(x_{i}, y_{i})$. In addition, each individual is assigned a weight $w_{i}$ that is a function of their position in society with respect to the welfare function’s goals. In other words, if we were using a utilitarian function, the weights would be independent of the their position as such a function isn’t concerned with the distribution of goods within a given population. On other hand, if we are concerned with distribution, then $w_{i}$ would be a function of their current income which is captured by $x_{i}$ feature set. With that, we can define basic NWP as the following,

\begin{equation}
NWP = \prod_{i=0}^{N} w_{i} U(x_{i},y_{i})\label{eq:1}
\end{equation}
given that $w_{i}$ and $U(x_{i}, y_{i})$ are non-negative

\subsection{Defining NWP for our binary classification problem}
Based on the above definition we can adapt the NWP for the post-processing phase of the fairness pipeline to modulate the machine prediction based on joint welfare and fairness constraints. We define prediction modulation as inflecting our prediction value based on utility differences as defined by NWP and our explicit policy goals. We will define both the utility differences and explicit policy goals in detail later in this section. First, for the purposes of defining our binary classification problem, we will consider one that uses a linear SVM classifier that seeks to minimize the accuracy loss that has been trained over a set of n points comprising of {$x_{i}, y_{i}$} where $x_{i}$ is a set of all the features of interest for the population and $y_{i}$ is the outcome to be modeled. Thus, the objective of the linear SVM classifier concerned with loss minimization can be described as,

\begin{equation*}
 \chi = \argmin \sum_{i=1}^{n} l(h(x_{i}),y_{i})
\end{equation*}

For a binary classification problem where $y_{i}$ $\epsilon$ {0,1} and we can simplify the loss function to consider only hinge loss for our described scenarios. To describe our binary classification problem, we will use terms institution and subject to describe the user of the prediction tool and the entity subject to those predictions respectively. We will initially use the example of a credit approval scenario to better translate the NWP from the economic domain to machine prediction. In such a case, a bank who serves as our institution is the creditor with a specified budget B and an unspecified policy constraint that describes how much from the B it can have loaned at any given time. If the subject, an individual i seeks to loan an amount, we will be interested in the set of features that describes their standing as a potential creditee in the eyes of the bank. This could use aspects like their current and revolving balance as well as credit evaluation metrics like FICO scores. In addition, the $x_{i}$ features could also include the protected attributes like race and gender. We can categorize two approaches -- race-blind and race-aware based on whether they utilize the protected attribute of race to inform their joint welfare and fairness constraints.
We thus adapt the NWP in Equation 1 to describe our generalized binary classification problem 

\begin{equation}
NWP_{binary} = w_{inst} U(x_{inst},y_{inst}) \prod_{i=0}^{N} w_{i} U(x_{i},y_{i})\label{eq:2}
\end{equation}

\emph{The Nash Welfare Product for a generalized binary classification problem becomes a simple product between the weighted utilities of the institution making the predictions and those individuals that are subject to the decision.
}

Note that the $x_{inst}$ and $x_{i}$ encompass the different features that contribute to the respective utilities of the institution and the subject. We can then define utility change as a central aspect in the prediction modulation step where if $NWP_{11}$ describes the scenario where a loan is given and repaid successfully and $NWP_{10}$ describes a default on the loan given, then we can clearly calculate the combined utility tradeoffs if the bank were to give a loan to an individual by 

\begin{equation*}
\Delta NWP_{1}  = NWP_{11} - NWP_{10}
\end{equation*}

Similarly, $\Delta NWP_{0}$ would describe the difference of the combined utility change for both the institution and the individual in the event the bank refuses to loan the desired amount to the individual and the individual would have repaid ($NWP_{01}$) and the same situation where the individual would have defaulted ($NWP_{00}$). With that, we get a way to easily compare simple utility changes for both the institution and the subject that can then inform the utility of making a particular decision in the context of binary classification.
\begin{equation}
U_{decision} = NWP_{1} - NWP_{0}\label{eq:3}
\end{equation}

Above, if $U_{decision}$ is non-negative then the utility of choosing y=1 exceeds that of y=0 and vice-versa. To this, we define our modulating function $\phi$ which is a function of the normalized hyperplane distance between the prediction point and the SVM hyperplane. We can then describe $\phi$(f(H)) as the modulating function that takes the hyperplane distance $\epsilon$ for a given prediction point defined by SVM classifier H and then modulates the prediction value itself based on a function of $U_{decision}$. Thus, we can define the modulating function $\phi$  as follows,

\begin{equation}
\phi(f(H)) = \omega(H(x_{i} , y_{i}),\epsilon, U_{decision})\label{eq:4}       
\end{equation}
\emph{The modulating function for a generalized prediction problem using a linear SVM can be defined as above by where $\omega$ is the combined function that alters the raw prediction value by utility costs described by $U_{decision}$ as a function of the hyperplane distance  which is taken obtained from the classifier H itself.
}

The above definition and its defining equation (6) underline an important point which combines our two-fold objective of loss minimization and societal utility maximization. Both $\epsilon$ and $U_{decision}$ directly control each other’s influence on the modulating function $\phi$. The extent to which they control each other is dependent on our next definition, namely explicit policy goals which we describe in the section below. 

\subsection{Making fairness versus welfare trade-offs explicit}

Central to our problem definition is the weights of each individual $w_{i}$ which as mentioned before describe the relational position of that individual in the society. If our explicit policy goal is \emph{welfare-oriented}, then the weights will be defined by a function $f_{welfare}$ which would seek to shift the weights accordingly to ensure even distribution. For instance, this would mean that a low-income individual will see their weights being increased to provide a boost to the potential NWP within the welfare condition. Thus, 

\begin{equation*}
f_{welfare} = f(NWP,x_{i},y_{i})
\end{equation*}

If our explicit policy goal is fairness-oriented, then the weights would be defined by $f_{fairness}$ that would be more concerned with minimizing disparate impact for an individual. This would mean that the prediction would be modulated in a way that would minimize harms for both parties with stakes in the prediction, the individual and institution. 
The motivation behind defining such explicit policy goals is to allow a more domain-specific approach to be taken by experts from our currently generalized hypothesis. Concretely, $f_{fairness}$ and $f_{welfare}$ may not always align but it is necessary to seek to make trade-offs between them explicit so policymakers as well as community members and other stakeholders can make informed decision about it. In case of the institution, $f_{welfare}$ would set their $w_{inst}$ to be a variable that would be dependent on the degree of distribution within the population, while $f_{fairness}$ would set $w_{inst}$ to be a constant at 0.5. 
Note that we do not consider futility because NWP as described in (2) would collapse both bank utility of maximizing their profit and the individual utility to preferably get their loan accepted into one equation. 

\subsection{Building a temporal feedback model}
As \cite{liu2018delayed}'s recent work explained, taking long-term impacts of the prediction is an important step forward from the current point in the fairness discourse if we wish to better understand and achieve our goals. In addition, we wish to expand upon our earlier critique we made about the existing fairness constraints being insufficient in removing “baked in” biases. Here, we expand upon \cite{liu2018delayed}’s work and describe our own version of temporal feedback model which takes into account the impact of the previous predictions and their associated outcomes. We then apply our modulation function  and explicit policy goals, $f_{welfare}$ and $f_{fairness}$ to counter the “baked in” biases that are embedded in the dataset. The shifting weights is the primary way in which the temporal feedback will inform the model over successive epochs. After each epoch, the model will evaluate its prediction errors, namely \emph{false positive} (FPR) and \emph{false negative rates} (FNR) and then balance them across groups by adjusting both individual and group weights in accordance with its policy goals. This can be seen as conceptually similar to basic reinforcement learning wherein the model adjusts its internal weights to penalize undesirable outcomes and shift likelihood towards desirable ones.

Thus, if the weight changes after an epoch are signified by a function $\tau$, then it can be defined as follows

\begin{equation}
\tau_{epoch} = \int_{i}^{N} \phi(f(H)) arg min (\chi_{fairness}, \chi_{welfare})\label{eq:5}
\end{equation}
\emph{where $\phi$  is our modulating function we defined earlier applied for the entire domain of values classified in a particular epoch while minimizing the mixed prediction loss $\chi$ based on welfare and fairness policy goals.
}

The above definition gives us a concrete way to extend our binary classification problem across multiple epochs. This allows us not just to model and understand the long-term effects of one-shot predictions as performed by \cite{liu2018delayed} but also provide us a way to analyze our hypothesis by simulating it for multiple epochs and understanding how its distributions shift based on the shifting weights which are described by $\tau_{epoch}$ .

\section{Results}

\subsection{Analysis on multi-epoch simulation}
To apply our temporal feedback model and empirically test the effects of NWP and modulating function, we chose two different scenarios. We first chose to simulate a loan-lending scenario by modeling its starting population data from UCI’s Adult Income dataset. We created a population test sample of n=100 that contained an equal number of White and African-American individuals. We normalized the income range between 100 and 1000 while adding a bit of gaussian noise to add variability to the rigid income categories in the UCI dataset. We set their initial weights to be a linear function of their income.

We performed simulation for M=6 epochs updating the weights after each epoch based on the function $\tau$ that was described earlier in Equation \ref{eq:5}. An epoch comprised of each individual agent seeking a loan that was picked from a normal distribution based on their current income and the bank making a decision based on the linear SVM classifier H with C=1.0 and gamma=0.5. The hyperplane distance for each prediction point gave its corresponding $\epsilon$ while the $U_{decision}$ was calculated based on net utility gain based on all possible scenarios. Then, as stated in Equation \ref{eq:4}, the prediction value was modulated by $\phi$ to provide with a decision of approval or rejection. This decision was then checked against the actual outcome, which itself was calculated by picking from a gaussian distribution with its mean set to the raw prediction, thus providing some variability in simulated outcomes. Based on how much the decisions deviated from the actual outcomes, we calculate the false positive and false negative rates and combined them into a prediction error rate. Based on the overall outcome, the individual agent as well as the bank gets the appropriate reward or penalty for each loan request. This marked the end of an epoch.

\begin{figure}
  \centering
	\includegraphics[width=0.5\textwidth]{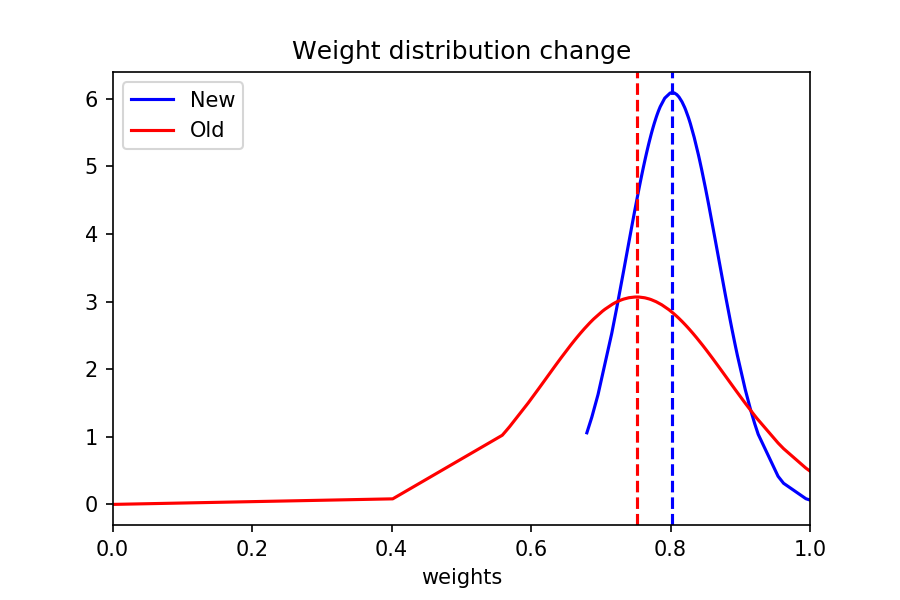}
  \caption{Plot showing the overall shift in distribution of individual weights after 6 rounds of simulation. Weights are a function of the individual income and thus their distribution is representative of the income distribution.}
  \label{fig:fig1}
\end{figure}

Before the next epoch, we updated the weights of every individual agent with $\tau_{epoch}$. Note that we keep the bank’s weights $w_{bank}$ to be a constant of 0.5 to give even weight to bank’s profit maximizing interest and society’s utility for the purposes of our test but it could also be used as a variable to satisfy different policy goals (for instance, $w_{bank}$ for a bank profit maximization scenario would be 1).

\begin{figure}
  \centering
	\includegraphics[width=0.5\textwidth]{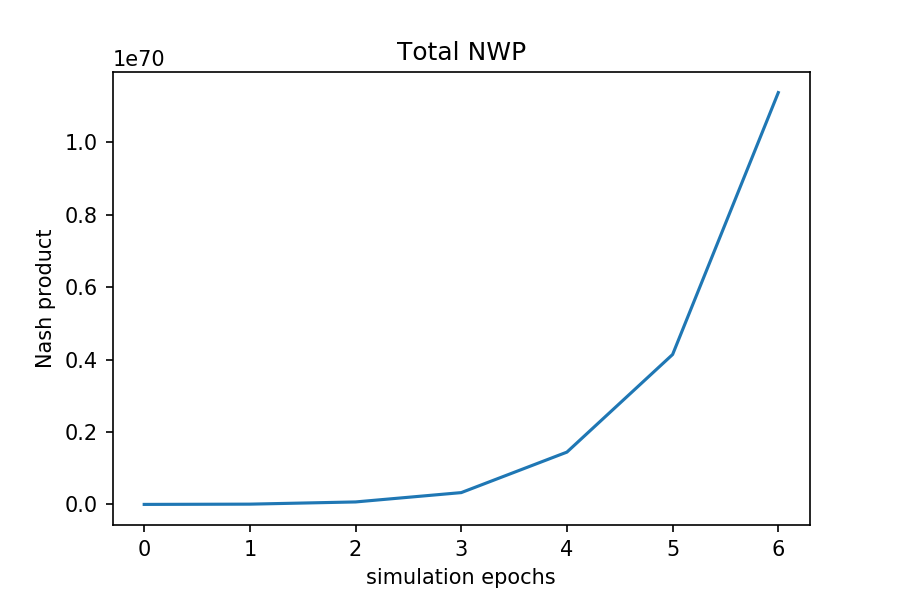}
  \caption{Plot showing the gradual increase of the Nash Welfare Product across successive simulation epochs}
  \label{fig:fig2}
\end{figure}

As Figure \ref{fig:fig2} shows, the Nash Welfare Product which collapses the utilities of both the institution (bank) using the prediction tool and the subject of the prediction (potential creditee) is shown to successively increase over 6 epochs of our simulation. A significant contribution in the increase in Nash Welfare Product is partly due to the increase in the mean of the weights and a shift in their overall distribution as illustrated in Figure \ref{fig:fig1}.

In addition, Figure \ref{fig:fig1}’s shift in distribution also underlines a similar shift in the income distribution as weights are themselves a direct function of an individual agent’s income. This is consistent with our hypothesis claiming that $f_{welfare}$ when utilized with NWP will seek to shift the overall distribution towards an equitable distribution. Figure 1 also underlines our point on utilizing NWP and associated methods proposed by us to self-correct the “baked in” biases of the dataset. Here, even though the initial weight and consequently income distributions differed based on recorded demographic differences in the UCI dataset, the differences tend to get evened out to an equilibrium after 6 epochs. For our analysis, we did not consider a \emph{race-blind} $f_{fairness}$ given that there is considerable literature as summarized by \cite{corbett2018measure} that goes into detail about whether excluding protected attributes might lead to unjustified disparate impact.

\begin{figure}
  \centering
	\includegraphics[width=0.5\textwidth]{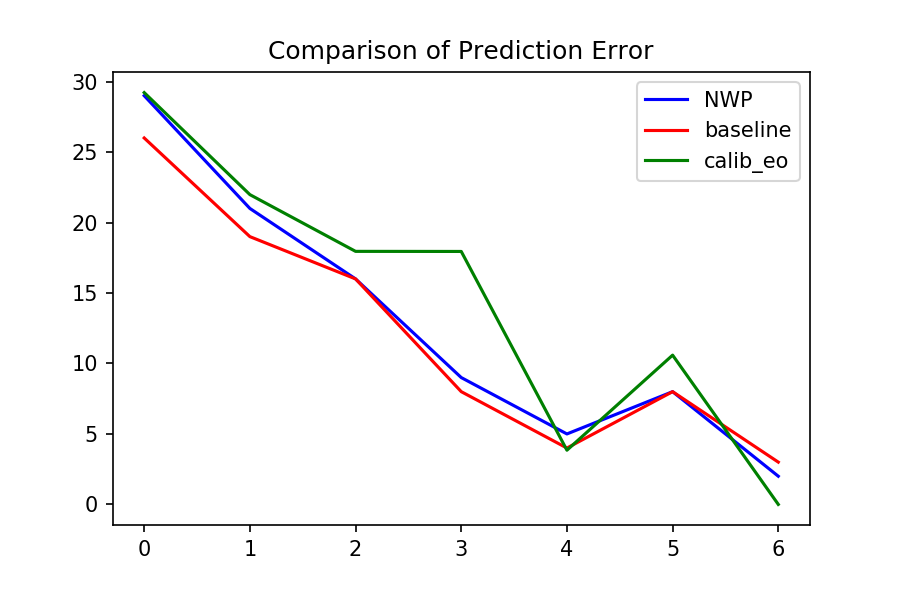}
  \caption{Plot showing a combined false positive and false negative rates of the linear SVM classifier across multiple cycles of simulation}
  \label{fig:fig3}
\end{figure}

In addition, the prediction error rates as defined before were traced over the 6 simulation epochs. We also applied an implementation of calibration as first shown in \cite{pleiss2017fairness} to compare our NWP-based weight correction compared with the calibrated equalized odds which made weighted adjustments to the false positive and false negative rates. As Figure \ref{fig:fig3} illustrates, the NWP-based weight correction we propose in this paper performs reasonably comparable to the calibrated equalized odds approach. The advantage that NWP brings is to provide corrective and distributive measures to move the “baked in” biases toward an equilibria as described by NWP which was shown in Figure \ref{fig:fig1}. 

\subsection{Test case analysis on ProPublica recidivism dataset}

We chose the ProPublica recidivism dataset for its familiarity reasons within the fairness research community. Since its original analysis by \cite{angwin2016machine} it has since then become a frequently used metric to compare fairness statistics. In particular, we were interested in proving our hypothesis of utilizing NWP that takes multiple features of both the institution and the subject into account when calculating utility. In this scenario, the institution is the society itself, or public safety in particular as the prediction tool is made by an the criminal justice system to ultimately serve for public good. The individual subject to the prediction is the person awaiting the pre-trial prediction. The features we used to model our societal and individual utility functions are age of the defendant, their race and prior counts. The rationale behind our choice can be explained by the following example. Take for instance, a young defendant with low prior counts who will have a potential high utility gain $U_{11}$ for both society and themselves where they are released and do not recidivate. They can contribute to society for longer in such a non-recidivating case, and the criminal justice provides them with a slight benefit of doubt to improve their future behavior. The rationale behind using race explicitly to model utilities is to counter the over-policing effect that has been showcased by numerous studies (Lum and Isaac 2016) before and we believe that the “baked in” biases as stated by Kleinberg et.al, 2018 cannot be corrected unless explicit policy decisions are made to both the actual over-policing issue and to the data that has its imprints. 

\begin{figure}
  \centering
	\includegraphics[width=0.5\textwidth]{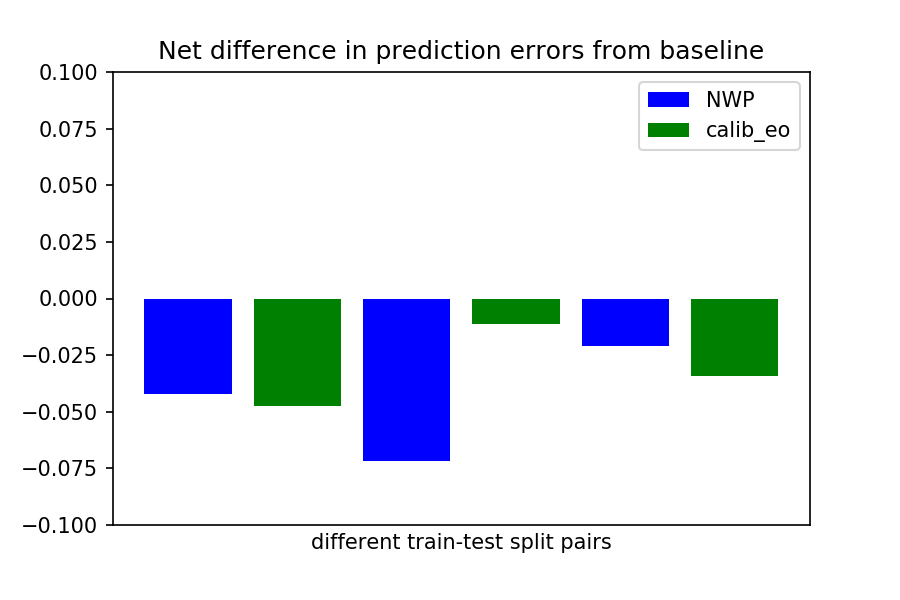}
  \caption{Comparison of NWP and calibrated equalized odds approaches in net difference from the baseline prediction error rates over three distinct different training-test data splits involving individuals with age less than 35 and prior counts less than 3}
  \label{fig:fig4}
\end{figure}

Thus, from the dataset we take a subset of population that is younger than or equal to 35 (age \textless = 35) and have a prior count of less than 3 (prior counts \textless 3). We then do a simple 70-30 training-testing split thrice to obtain 3 distinct variations of the subpopulation. We then analyze the baseline prediction error rates characterized by false positive and false negative rates and then compare it to just a single epoch (M=1) of NWP and calibrated equalized odds (as adapted from \cite{pleiss2017fairness}) separately. As Figure \ref{fig:fig4} illustrates, the NWP performs comparably to slightly better than calibrated equalized odds at reducing the overall prediction error rates compared to the baseline. We believe this is primarily due to incorporating $U_{decision}$ by modeling utility changes in different scenario for both the institution (public safety) and the subject of the prediction (defendant). 

\section{Discussion}

As illustrated above in both the multi-epoch simulation of the loan-lending scenario and the test case analysis on a subgroup of ProPublica dataset, underlines our hypothesis that expanding the scope of fairness and subsequent utility calculations to include the subject of the prediction results in fairer outcomes. This goes beyond the application of fairness constraints on the one-shot prediction model that is currently dominant in the literature. As showcased in the results of our simulation with Figure \ref{fig:fig1}, NWP performs a shift of income distribution (or risk distribution, as it is often generalized in the literature) while maintaining comparable prediction error rates to the dominant methods in the literature like calibrated equalized odds.
A core reason for this success is similar to why NWP works so well in other domains of game theory and multi-agent allocation. Namely, the multiplicative nature of NWP collapses the utility of individuals and the institution into a common equation, with the underlying differences in distributions getting eventually pushed towards an equilibrium, with the rate of equilibrium proportional to $f_{welfare}$ > $f_{fairness}$. By modeling the weights which influence the utility of the stakeholders we get a better way to control them with  epoch  shifting the weights according to how $f_{fairness}$ and $f_{welfare}$ are specified by the policy goals. The negotiation between $f_{welfare}$ and $f_{fairness}$ also gets to the heart of the infra-marginality critique made by \cite{corbett2018measure} and by solving for $\tau_{epoch}$, we get a self-corrective mechanism to undo biases in the dataset by moving it towards an equitable risk distribution over multiple epochs as shown in Figure \ref{fig:fig1}. 

The results from our test case analysis on the younger subpopulation with low prior counts from the ProPublica recidivism dataset further emphasizes our point on the importance of expanding the scope of utility. By modeling utility to include features like age, prior counts and race we showcase that the predictive error rates can be further improved. By using that and even going even beyond that by using $\tau_{epoch}$ to determine shifts in individual and subgroup weights we provide correct mechanisms for biases that exist in the dataset and the practices like over-policing which continue to shape such data. As stated before, we cannot achieve fair outcomes without fully addressing the scope of unfairness that is already embedded within the data.

\section{Future Work}

Beyond this paper, the biggest challenge continues to lie in expanding thinking about fairness beyond the currently dominant discussion.  Several other promising avenues of emerging research have been outlined earlier in the paper (see Recent Work). Apart from that, we believe moving beyond the existing focus on one-shot prediction is important. In addition, research on improving techniques of causal inference and machine interpretability (\cite{doshi2017towards}) remain an important challenge.
The challenges posed by modeling utilities based on complex, interconnected features isn’t a new problem and has been a subject of research in fields like economics which will continue to be an important resource for future research within the domain of algorithmic prediction. Particularly because we believe the current state of research has reached a point where complex tweaks and cost adjustments are mostly just shuffling the underlying problem under the rug. Namely, the existing biases embedded in the dataset cannot be simply corrected in the vacuum of a one-shot prediction and then be forgotten. They need a more corrective mechanism we propose in this paper and beyond that obvious policy-making changes to contextually enhance how these technologies are applied in their respective domains. As \cite{corbett2018measure} pointed out, if the limitations of existing fairness constraints is partly due to the varying risk distributions, then we believe that shifting weights $\tau_{epoch}$  is one possible way we can try to eventually seek to minimize differences in distributions. Suggestions made by \cite{hu2018welfare} in their Social Planner Problem suggest an alternative approach along this direction as well.
Our central belief for future research remains to have an open, interdisciplinary approach to grapple with a similarly complex issue. Bringing in literature and expertise from other fields will be essential to not just solving domain-specific problems but also solving computer-specific problems. 

\section{Conclusion}

In this paper, we proposed a critique of the existing way of thinking about fairness and sought to reconceptualize it by proposing a novel method. The methods we proposed expands how we think about fairness and calculate associated utility of possible outcomes by combining both the institution and subject utilities and modulating it based on how we seek to address the “baked in” biases of the dataset. The two-step method was then tested on a multi-epoch simulation modeled of the UCI Adult Income dataset and was shown to shift the initial distribution towards an equitable distribution while maintaining comparable prediction error rates to existing fairness methods like calibrated equalized odds. We also showcased the importance of modeling utility based on carefully chosen attributes with our test case analysis of the ProPublica dataset.
This further proves that our proposed method of combining institution and subject utility with NWP and modulating prediction results in a fairer outcome not just for the current epoch but also provides corrective measures to overcome its baked in biases.

%\bibliographystyle{apalike}
%\bibliography{references}

\end{document}